\documentclass[aps,prc,twocolumn]{revtex4}
\usepackage{graphicx,amssymb}

\begin{document}

\title{Local Causality and Completeness:  Bell \emph{vs.} Jarrett}
\author{Travis Norsen}
\affiliation{Marlboro College \\ Marlboro, VT  05344 \\ norsen@marlboro.edu}

\date{August 14, 2008}

\begin{abstract}
J.S. Bell believed that his famous theorem entailed a deep and
troubling conflict between the empirically verified predictions of
quantum theory and the notion of local causality that is motivated by
relativity theory.  Yet many physicists continue to accept, usually on 
the reports of textbook writers and other commentators, that Bell's
own view was wrong, and that, in fact, the theorem only brings out a
conflict with determinism or the hidden-variables program or realism
or some other such principle that (unlike local causality), allegedly,
nobody should have believed anyway.  (Moreover, typically such beliefs
arise without the person in question even being aware that the view
they are accepting differs so radically from Bell's own.)
Here we try to shed some light
on the situation by focusing on the concept of local causality that is
the heart of Bell's theorem, and, in particular, by contrasting Bell's
own understanding with the analysis of Jon Jarrett which has been the
most influential source, in recent decades, for the kinds of claims
mentioned previously.  We point out a crucial difference between 
Jarrett's and Bell's own understanding of Bell's formulation of 
local causality, which turns out to be the basis for the erroneous 
claim, made by Jarrett and many others, that Bell misunderstood 
the implications of his own theorem. 
\end{abstract}

\maketitle

\section{Introduction}

In 1964, J.S. Bell proved the result now known as Bell's
Theorem:  any physical theory of a certain
type must make predictions (for a certain class of experiment) which
respect a so-called Bell Inequality.  
\footnote{John S. Bell, \emph{Speakable and Unspeakable in Quantum
    Mechanics}, 2nd ed., Cambridge University Press, 2004.  Subsequent
  references to Bell's writings in the text will be given in-line with
  the year of the referenced paper and page numbers from the book.}
Quantum Mechanics (QM) predicts
violations of the inequality, and subsequent
experiments establish convincingly (though not without loopholes) that
the Quantum Mechanical predictions are correct -- i.e., the
experiments establish that 
the type of theory Bell showed must respect the Inequality, cannot be
empirically viable, i.e., cannot be true.  
\footnote{For a review of recent experiments and associated loopholes,
  see, e.g., Abner Shimony, ``Bell's Theorem'', \emph{The Stanford
  Encyclopedia of Philosophy} (Fall 2006 Edition), Edward N. Zalta
(ed.), URL = http://plato.stanford.edu/archives/fall2006/entries/bell-theorem/}
But the
question (which has given rise to an enormous literature) remains:
what type of theory is it, exactly, that Bell's Theorem (combined with
the associated experiments) refutes?

Bell's own view, expressed already in the opening lines of his 1964
paper and subsequently clarified and defended in virtually all of his
later writings, was that ``It is the requirement of locality ...
that creates the essential difficulty.''  (Bell, 1964, p. 14)
By ``locality'' Bell here means 
the prohibition, usually taken to be an implication of
special relativity (SR), of super-luminal (faster than light) 
causation.  Bell thus 
took his own theorem to establish a troubling conflict between (the
empirically verified predictions of) QM (i.e.,
between \emph{experiment}) and SR:
\begin{quote}
``For me then this is the real problem with quantum theory:  the
apparently essential conflict between any sharp formulation and
fundamental relativity.  That is to say, we have an apparent
incompatibility, at the deepest level, between the two fundamental
pillars of contemporary theory...'' (Bell, 1984, p. 172)
\end{quote}
Most practicing physicists, however, (and most philosophers of
physics) have disagreed with Bell and continued to believe in the
unproblematic consistency of QM and SR.  Where do they think Bell went
wrong?  

One can divide reasons for disagreement (with Bell's own interpretation
of the significance of his theorem) into two classes.  First, there are
those who assert that the derivation of a Bell Inequality relies not
just on the premise of locality, but on some additional premises
as well.  The usual suspects here include Realism, Hidden Variables,
Determinism, and Counter-Factual-Definiteness.  (Note that the items
on this list are highly overlapping, and often commentators use them
interchangeably.)  The idea is then that, since it is only the
\emph{conjunction} of locality with some other premise which is in conflict
with experiment, and since locality is so strongly motivated by SR, we
should reject the other premise.  Hence the widespread reports that
Bell's theorem finally refutes the hidden variables program, the
principle of determinism, the philosophical notion of realism, etc.
\footnote{See, for example:  N. David Mermin, ``What is quantum
  mechanics trying to tell us?'' \emph{AmJPhys}, {\bf{66}}(9),
  September 1998, pg 753-767; Marek Zukowski, ``On the paradoxical
  book of Bell,'' \emph{Stud. Hist. Phil. Mod. Phys.}, {\bf{36}}
  (2005) 566-575; A. Zeilinger, ``The message of the quantum,''
  \emph{Nature} {\bf{438}}, 743 (8 December, 2005); Daniel Styer,
  \emph{The Strange World of Quantum Mechanics} (page 42), Cambridge,
  2000; George Greenstein and Arthur Zajonc, \emph{The Quantum
    Challenge} (Second Edition), Jones and Bartlett Publishers,
  Sudbury, Massachusetts, 2006; John Townsend, \emph{A Modern Approach
  to Quantum Mechanics}, McGraw-Hill, 1992; Herbert Kroemer
\emph{Quantum Mechanics}, Prentice Hall, New Jersey, 1994; Richard
Liboff, \emph{Introductory Quantum Mechanics} (2nd edition),
Addison-Wesley, Reading, Massachusetts, 1992}

Here is how Bell responded to this first class of disagreement:
\begin{quote}
``My own first paper on this subject ... starts with a
summary of the EPR argument \emph{from locality to} deterministic
hidden variables.  But the commentators have almost universally
reported that it begins with deterministic hidden variables.''
(Bell, 1981, p. 157)
\end{quote}
Here
(a footnote, but also in the main text of the article in question)
Bell goes out of his way to stress the overall logical structure of
his \emph{two-part} argument:  first, an argument \emph{from} locality and
certain predictions of QM (namely, perfect anti-correlation for
parallel spin measurements on a pair of spin 1/2 particles in the spin
singlet state) \emph{to} the existence of deterministic local hidden
variables; and then second, from such variables to the inequality,
i.e., to a disagreement with certain other predictions of QM.  This
whole first class of disagreement with Bell, then, rests on a simple
confusion about Bell's argument. 
\footnote{For further discussion, see any of Bell's papers and, e.g.:
  Tim Maudlin, \emph{Quantum Non-Locality and Relativity} (Second
  Edition), Blackwell, Malden, Massachusetts, 2002;  Travis Norsen,
  ``Bell Locality and the Nonlocal Character of Nature,''
  \emph{Found. Phys. Lett.}, {\bf{19}}(7), 633-655 (Dec. 2006)}

The more interesting and more subtle second class of disagreement 
includes those who accept that the empirically-violated Bell
inequality can be derived from Bell's locality condition alone, but
who argue that this locality condition is \emph{too strong}, i.e., 
that it smuggles in some extra requirements beyond those
minimally necessary to respect SR's prohibition on superluminal
causation.  At the head of this class is Jon Jarrett, whose 1983 
PhD thesis and subsequent 1984 paper 
\footnote{Jon Jarrett, ``On the Physical Significance of the Locality
  Conditions in the Bell Arguments,'' \emph{Nous} {\bf{18}} (1984) 569-589}
argued that Bell's own local causality condition (which Jarrett calls 
``strong locality'') is logically equivalent to
the conjunction of two subsidiary conditions, which Jarrett described
respectively as ``locality'' and ``completeness.'' 

Roughly speaking, Jarrett's ``locality'' is the requirement that the
outcome of a measurement on one particle be independent of the type of
measurement performed (at spacelike separation) on a second particle
(which, in the interesting sorts of cases, is described by QM as being
entangled with the first particle).  Jarrett's ``completeness'' on the
other hand requires the outcome of the first measurement to be
independent of the outcome of the second, spacelike separated
measurement.  He then argues that a violation of ``locality'' would
entail the ability to send superluminal signals (Alice could learn
something about the setting of Bob's measurement device by examining
the outcome of her own experiment) which would allow Bob to transmit
a signal across a spacelike interval, something clearly forbidden
by SR.  By contrast, a violation of ``completeness'' indicates no
conflict with SR.  Thus, Jarrett
argues, in the face of the empirical data conflicting with ``strong
locality'' we may reject ``completeness'' and thereby achieve -- contra
Bell -- a kind of Peaceful Coexistence between QM and SR.  
\footnote{``Peaceful Coexistence'' is Abner Shimony's term:  
  ``Metaphysical problems in the foundations of quantum mechanics,''
  \emph{International Philosophical Quarterly} 18, 3-17.}

Jarrett's project has been widely hailed and widely discussed.  It was
the immediate stimulus for almost everything in the ``Bell
literature'' for about a decade after its appearance, and continues to
set a broadly influential context for much ongoing work in this area.
\footnote{See, for example, M.L.G. Redhead, \emph{Incompleteness,
    Nonlocality, and Realism: A Prolegomenon to the Philosophy of
    Quantum Mechanics} Oxford, 1987; J. Cushing and E. McMullin, eds.,
  \emph{Philosophical Consequences of Quantum Theory}, Notre Dame,
  1989 (see especially the contributions of Paul Teller and Don
  Howard); Brandon Fogel, ``Formalizing the separability condition in
  Bell's theorem,'' \emph{Stud. Hist. Phil. Mod. Phys.}, 38 (2007),
  920-937} 

The purpose of the present paper is to critically assess
Jarrett's analysis and
the conclusions it led him to, by comparing his project side-by-side
with Bell's own discussions of the relevant issues.  In particular, I
will claim that Jarrett simply missed a crucial aspect (having
something to do with ``completeness'') of Bell's
formulation of local causality; this turns out to be the heart of the
thinking behind Jarrett's (\emph{prima facie} rather puzzling) terminology
for his two sub-conditions, as well as his central claim that
violations of his ``completeness'' criterion indicate no conflict with
special relativistic local causality.  The main conclusion is thus
that, contrary to Jarrett and his followers, Bell's own local 
causality criterion is in no sense ``too strong.''  And this of course
undermines the attempt to establish the Peaceful Coexistence of QM and
SR, i.e., it supports Bell's own interpretation of the meaning of his
theorem. 

The following two
sections present, respectively, Bell's own (final and most careful)
formulation of the locality premise, and then Jarrett's analysis.  
Section IV includes some comparative discussion, highlighting
especially the relation of Jarrett's thinking to the EPR argument.  A
brief final section then summarizes and concludes.

Before launching into this, however, it is appropriate to briefly
survey earlier criticisms of Jarrett's analysis of Bell's locality
concept.  To my knowledge, the fullest critical discussion of
Jarrett's project is in Maudlin's book. \footnote{\emph{Op cit., 
pp. 93-98}}  
Maudlin's main points \emph{vis-a-vis} Jarrett are as follows:  
(a) Jarrett's identification of his ``locality'' sub-condition with
the prohibition on superluminal signals is wrong; (b) Jarrett's
identification of superluminal signaling with superluminal causation
is wrong; (c) Jarrett's claim that a violation of his ``completeness''
condition does not entail any nonlocal causation is wrong.
\footnote{Similar points are made in Jeremy Butterfield, ``Bell's
  Theorem: What it Takes'' \emph{Brit. J. Phil. Sci.}, {\bf{43}}
  (1992) 41-83.  Butterfield, however, (unlike Maudlin) ends up
  accepting both the validity/meaningfulness of Jarrett's analysis and
  Jarrett's ultimate conclusion about Peaceful Coexistence, 
  even though he disagrees with some of Jarrett's arguments.  Since I
  disagree with Butterfield's conclusions, I find him a less
  convincing overall critic of Jarrett's project than Maudlin.  My
  objection to Butterfield's proposed route to Peaceful Coexistence --
  which, like the present discussion of Jarrett, involves the claim 
  that the commentator has missed or misunderstood a crucial element
  of Bell's concept of local causality -- will be presented elsewhere.}

The arguments for (a) and (b) are made clearly and compellingly by
Maudlin, who thus really demolishes Jarrett's erroneous identification
of his ``locality'' with the relevant requirements of SR.  But the
case for (c) is made only indirectly, essentially by dismissing
Jarrett and re-asserting Bell's claim to the contrary.  So, to be a
bit more precise about the goal of the present paper, the aim is to
fill this gap by exploring in detail how Jarrett's ``completeness''
condition relates to Bell's local causality criterion and how Jarrett's
misunderstanding of the latter led him to the various erroneous
conclusions.  

I should note at the outset, however, that the view to be presented
here as Jarrett's is almost certainly a bit misleading as to his 
(or those I consider his followers') fully-considered views.  Jarrett
does actually say all the things I attribute to him, but my gloss will
perhaps minimize the extent to which Jarrett (I would argue,
inconsistently) also acknowledges points in conflict with the views I
will attribute to him.  It is probably best, therefore, to understand
the ``Jarrett'' discussed here as a rhetorically clarifying construct,
which may or may not correspond to the views of the actual Jon
Jarrett.

\section{Bell's Concept of Local Causality}

\begin{figure}[t]
\begin{center}
\includegraphics[width=3.3in,clip]{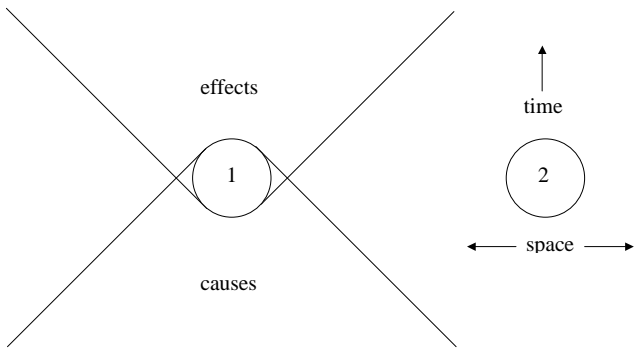}
\end{center}
\caption{
\label{fig1}
``Space-time location of causes and effects of events in region 1.''
(Figure and caption are from Bell, 1990, p. 239.)
}
\end{figure}

Bell's fullest and evidently most-considered discussion of local
causality occurs in his last published paper, \emph{La nouvelle
cuisine} (1990, 232-248).  We will here essentially follow that discussion,
supplementing it occasionally with things from his earlier papers.

Bell first introduces what he calls the ``Principle of local
causality'' as follows:  
``The direct causes (and effects) of events are near by, and even the
indirect causes (and effects) are no further away than permitted by
the velocity of light.''
Then, referencing what has been reproduced here as Figure 1, Bell
elaborates:  ``Thus, for events in a space-time region 1 ... we would
look for causes in the backward light cone, and for effects in the
future light cone.  In a region like 2, space-like separated from 1,
we would seek neither causes nor effects of events in 1.  Of course
this does not mean that events in 1 and 2 might not be correlated...''  
(1990, p. 239)

After remarking that this formulation ``is not yet sufficiently sharp
and clean for mathematics,'' Bell then proposes the following 
version, referencing what has been reproduced here as Figure 2:
\begin{quote}
``A theory will be said to be locally causal if the probabilities
attached to values of local beables in a space-time region 1 are
unaltered by specification of values of local beables in a space-like
separated region 2, when what happens in the backward light cone of 1
is already sufficiently specified, for example by a full specification
of local beables in a space-time region 3...'' (1990, 239-40)
\end{quote}
Although Bell doesn't immediately formulate this mathematically (which
is curious, since he has just advertised it as a formulation which
\emph{is} ``sufficiently sharp and clean for mathematics''), we may do
so in a way that is clearly (as evidenced by what comes later in the
paper) what he had in mind:
\begin{equation}
P(b_1 | B_3, b_2) = P(b_1 | B_3).
\label{eq:loc}
\end{equation}
Here $b_i$ refers to some beable (or more precisely, its value) 
in region $i$, and $B_i$ refers to a
``full specification'' of beables in region $i$.  This simply asserts 
mathematically what Bell states in the caption of his
accompanying figure:  ``full specification of [beables] in 3 makes 
events in 2 irrelevant for predictions about 1.''  
Note that Bell here uses the term (which he had earlier coined)
``beable'' (rhymes with ``agreeable'')
to denote whatever is posited, by the candidate theory in question, 
to be \emph{physically real}:  
\begin{quote}
``The beables of the theory are those elements which might
correspond to elements of reality, to things which exist.  Their 
existence does not depend on `observation'.  Indeed observation
and observers must be made out of beables.'' (1984, p. 174)
\end{quote}
For further discussion of ``beables'' see Bell's \emph{The Theory of
  Local Beables} (1975, pages 52-3), \emph{Beables for quantum field
  theory} (1984, pages 174-6), and \emph{La nouvelle cuisine} (1990,
  pages 234-5).

\begin{figure}[t]
\begin{center}
\includegraphics[width=3.3in,clip]{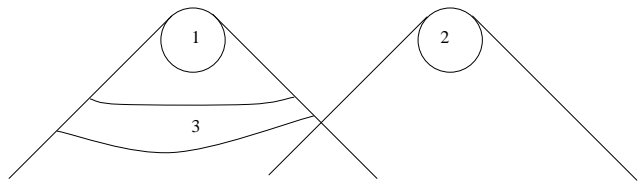}
\end{center}
\caption{
\label{fig2}
``Full specification of what happens in 3 makes events in 2 irrelevant
for predictions about 1 in a locally causal theory.''  
(Figure and caption are from Bell, 1990, p. 240.)
}
\end{figure}

Bell then adds the following clarificatory remarks:
\begin{quote}
``It is important that region 3 completely shields off from 1 the
overlap of the backward light cones of 1 and 2.  And it is important
that events in 3 be specified completely.  Otherwise the traces in
region 2 of causes of events in 1 could well supplement whatever else
was being used for calculating probabilities about 1.  The hypothesis
is that any such information about 2 becomes redundant when 3 is
specified completely.''  (1990, p. 240)
\end{quote}
It will be crucial to understand these remarks, so we shall briefly
elaborate.  

First, suppose that the region labeled $3$ in Figure 2
sliced across the backwards light cone of 1 at an earlier time, such
that it failed to ``completely shield off from 1 the overlap of the
backward light cones of 1 and 2.''  See, for example, the region 
labeled $3^*$ in Figure 3.  Why then would a violation of
Equation \ref{eq:loc} (but with $3 \rightarrow 3^*$) 
fail to necessarily indicate the presence of 
nonlocal causation?  Suppose we are dealing with a non-deterministic
(i.e., irreducibly stochastic, genuinely chancy) theory.  And suppose
that some event (a beable we shall call ``X'')  comes into existence
in the future of this
region $3^*$, such that it lies in the overlapping
backwards light cones of $1$ and $2$.  By assumption, the theory in
question did not allow the prediction of this beable on the basis of a
full specification of beables in $3^*$.  Yet once X comes into existence,
it can (in a way that is perfectly consistent with local causality)
influence events in both 1 and 2.  And there is therefore the
possibility that specification of events from 2 could allow one to infer
something about X from which one could in turn infer \emph{more}
about goings-on in 1 than one could have inferred originally from just
the full specification of beables in $3^*$.  In other words (Bell's): ``the
traces in region 2 of causes [such as our X] of events in 1
could well supplement whatever else was being used for calculating
probabilities about 1.''  This mechanism for producing violations of
Equation \ref{eq:loc} without any superluminal causation, however,   
will clearly not be available so long as ``region 3 completely shields
off from 1 the overlap of the backward light cones of 1 and 2.''

\begin{figure}[t]
\begin{center}
\includegraphics[width=3.3in,clip]{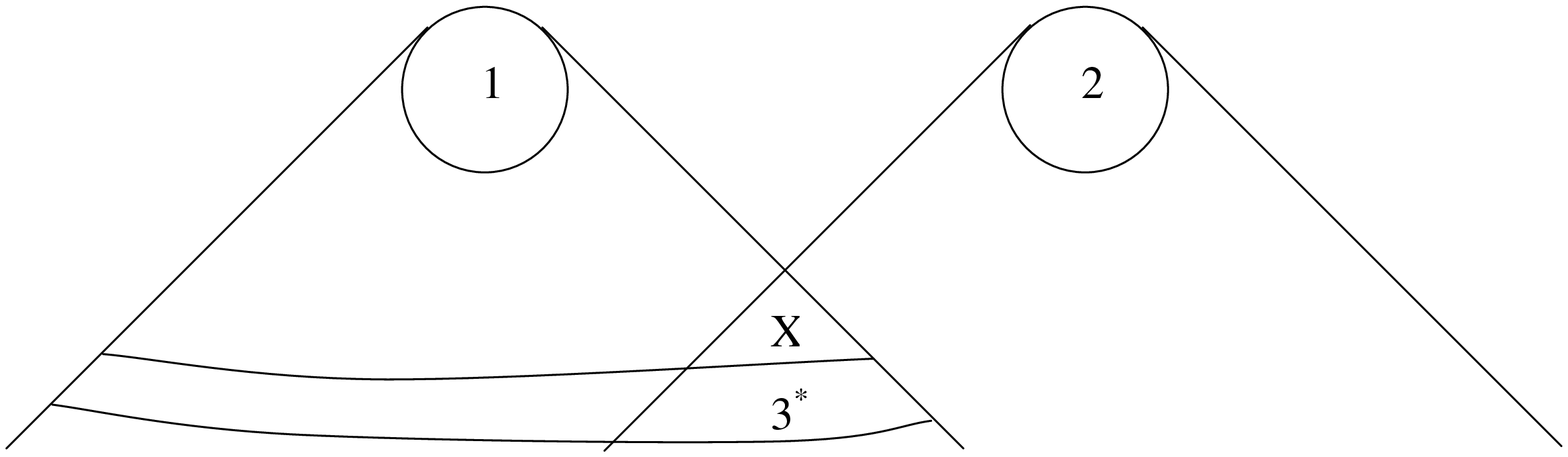}
\end{center}
\caption{
\label{fig3}
Similar to Figure 2, except that region $3^*$ (unlike region 3 of
Figure 2) fails to shield off region 1 from the overlapping backward
light cones of regions 1 and 2.  Thus, (following the language of
Figure 2's caption) even full specification of  what happens in $3^*$ 
\emph{does not necessarily make} events in 2 irrelevant for 
predictions about 1 in a locally causal theory.  
}
\end{figure}

Bell's other clarification is also crucial.  Suppose that events in
region 3 of Figure 2 are \emph{not} specified completely.  We may
denote such an incomplete description by $\bar{B}_3$.  Then does a
violation of Equation \ref{eq:loc} (but with $B_3 \rightarrow \bar{B}_3$)
necessarily imply the existence of any nonlocal causation?  No, for it
would then be possible that some event X (again in the overlapping past
light cones of 1 and 2) influences both 1 and 2 such that
information about 2 could tell us something about X which in turn
could tell us something about 1 which we couldn't infer from $\bar{B}_3$.
We only need stipulate that the beables in region 3 which ``carry''
the causal influence from X to 1 are (among) those omitted by $\bar{B}_3$.
But since there is, by definition, no such omission in the \emph{complete}
specification $B_3$, this eventuality cannot arise, and a violation of
Equation \ref{eq:loc} \emph{must} indicate the existence of some
nonlocal causation, i.e., causal influences not respecting Bell's
original ``Principle of local causality'' (as displayed in Figure 1).  
\footnote{Note that Bell stresses the need for a \emph{complete}
specification of beables in the relevant space-time region already
in his 1975 paper \emph{The theory of local beables}:  
``Now my intuitive notion of local causality is
that events in 2 should not be `causes' of events in 1, and vice
versa.  But this does not mean that the two sets of events should be
uncorrelated, for they could have common causes in the overlap of
their backward light cones.  It is perfectly intelligible then that if
[$B_3$] in [our Equation \ref{eq:loc}] 
does not contain a complete record of events ... it
can be usefully supplemented by information from region 2.  So in
general it is expected that [$P(b_1|B_3,b_2) \ne P(b_1|B_3)$.]
However, in the particular case that [$B_3$] contains already a
\emph{complete} specification of beables ... supplementary information
from region 2 could reasonably be expected to be redundant.'' 
(1975, p. 54)  Emphasis in original.  
This is especially relevant since we will eventually criticize
Jarrett for failing to appreciate (in his 1984 paper) this particular 
aspect.  So it shouldn't be thought that we are criticizing him for 
something Bell only understood and clarified later.  It is worth
noting, however, that there are some interesting differences between
Bell's 1975 and 1990 formulations of local causality; these will be
explored elsewhere, though, since they do not bear on Jarrett's analysis.}

It is worth noting that we cannot necessarily infer, from a violation
of Equation \ref{eq:loc}, that $b_2$ exerts any direct or indirect
causal influence on $b_1$.  It might be, for example,
that a violation of Equation \ref{eq:loc} is produced by some X-like
event lying in the future of region 3, which causally
influences both 1 and 2.  But in order to exert a \emph{local} causal
influence on 1, such an X would have to lie \emph{outside} the past light
cone of 2 -- and vice versa.  What is ensured by a violation of Equation
\ref{eq:loc} is thus only that \emph{some} violation of local
causality (as sketched in Figure 1) is being posited somewhere by the
theory in question.  Whether something in 2 is exerting a causal
influence on 1 (or vice versa, or neither) the mere violation of
Equation \ref{eq:loc} doesn't permit us to say.  
\footnote{This is why, despite being an improvement over Jarrett's
  terminology for the two sub-conditions to be discussed in Section
  III, Abner Shimony's terminology (``parameter independence'' for
  what Jarrett calls ``locality'' and ``outcome independence'' for
  what Jarrett calls ``completeness'') is also dubious.  For the
  terminology implies that a violation of one of the conditions
  entails that the event in question causally \emph{depends} on the distant
  ``parameter'' or ``outcome'' respectively.  But this need not be the
  case.}

Note that
everything in the above discussion refers to some particular candidate
physical theory.  For example, there is a tendency for misplaced
skepticism to arise from Bell's use of the concept of ``beables'' in 
the formulation of local causality.  This term strikes the ears of
those influenced by orthodox quantum philosophy as having a
metaphysical character and/or possibly committing one (already, in the
very definition of what it means for a theory to respect relativistic
local causality) to something unorthodox like ``realism'' or ``hidden
variables.''  Such concerns, however, are based on the failure to 
appreciate that
the concept ``beable'' is theory-relative.  ``Beable'' refers not to
what \emph{is} physically real, but to what some candidate theory
\emph{posits} as being physically real.  Bell writes:
``I use the term
`beable' rather than some more committed term like `being' or `beer'
to recall the essentially tentative nature of any physical theory.
Such a theory is at best a \emph{candidate} for the description of
nature.  Terms like `being', `beer', `existent', etc., would seem to
me lacking in humility.  In fact `beable' is short for `maybe-able'.''
(1984, p. 174)

Similar considerations apply to the notion of ``completeness'' that
is, as stressed above, essential to Bell's
formulation.   A  complete specification of
beables in some spacetime region simply means a specification of
everything (relevant) that is posited by the candidate theory in
question.  There is no presumption that such a full specification
actually correspond to what \emph{really exists} in the relevant
spacetime region, i.e., no presumption that the candidate theory in
question is \emph{true}.
And the same goes for the probabilities in Equation
\ref{eq:loc} that Bell's locality criterion is formulated in terms
of.  These should be read not as empirical frequencies or subjective
measures of expectation, but as the fundamental dynamical
probabilities described by the candidate theory in question (which we
assume, without loss of generality, to be irreducibly stochastic).
\footnote{Determinism is simply a special case in which all
  probabilities are either zero or unity.}

Since all the crucial aspects of Bell's formulation of locality are
thus meaningful only relative to some candidate theory, it is perhaps 
puzzling how Bell thought we could
say anything about the locally causal character of Nature.  Wouldn't
the locality condition only allow us to assess the local character of
candidate theories?  It is important to understand that the
answer is essentially (at least initially):  Yes!  Indeed, note that
Bell begins the formulation with ``A \emph{theory} will be said to be 
locally causal if...'' (emphasis added).   Let us state it openly and
explicitly:  Bell's locality criterion is a way of distinguishing
local theories from nonlocal ones:  
\begin{quote}
``I would insist here on the distinction
between analyzing various physical theories, on the one hand, and
philosophising about the unique real world on the other hand.  In this
matter of causality it is a great inconvenience that the real world is
given to us once only.  We cannot know [by looking] what would have 
happened if something had been different.  We cannot repeat an
experiment changing just one variable; the hands of the clock will
have moved, and the moons of Jupiter.   Physical 
theories are more amenable in this respect.  We can \emph{calculate}
the consequences of changing free elements in a theory, be they only
initial conditions, and so can explore the causal 
structure of the theory.  I insist that [my concept of local causality]
is primarily an analysis of certain kinds of physical theory.''
(1977, p. 101)
\end{quote}
How then did Bell think we could end up saying something interesting
about Nature?  That is precisely the beauty of Bell's theorem, which
shows that no theory respecting the locality condition (no matter what
other properties it may or may not have -- e.g., hidden variables or
only the non-hidden sort, deterministic or stochastic, particles or
fields or both or neither, etc.) can agree with the
empirically-verified QM predictions for certain types of experiment.
That is (and leaving aside the various experimental loopholes), no 
locally causal theory in Bell's sense
can agree with experiment, can be empirically viable, can be
true.  Which means the true theory (whatever it might be) necessarily
\emph{violates} Bell's locality condition.  Nature is not locally
causal.  
\footnote{This sometimes comes as a shock to adherents of orthodox
quantum theory, who are used to thinking of their own theory --
especially in its allegedly relativistic variants -- as perfectly
consistent with SR.  But the nonlocality of orthodox QM is quite
obvious, if one knows where to look.  The key here is that the
theory is not defined exclusively by the Schr\"odinger (or
equivalent) dynamical equation, but also by some version of a
collapse postulate.  And this latter is explicitly nonlocal.
Indeed, orthodox (collapse) QM is even \emph{more} nonlocal than
certain alternative theories, like Bohmian Mechanics, which are 
often maligned precisely for displaying an obvious kind of nonlocality.
The simplest type
of example which suffices to make this point is the ``Einstein's
Boxes'' scenario.  (See Travis Norsen, \emph{AmJPhys} 73(2), Feb. 2005,
pages 164-176.)  Bell explains beautifully how
this scenario manifests the nonlocal causation inherent in orthodox
QM:  ``Suppose ... we have a radioactive nucleus which can emit a
single $\alpha$-particle, surrounded at a considerable distance by
$\alpha$-particle counters.  So long as it is not specified that
some \emph{other} counter registers, there is a chance for a
particular counter that \emph{it} registers.  But if it is specified
that some other counter does register, even in a region of
space-time outside the relevant backward light cone, the chance that
the given counter registers is zero.  We simply do not have
[Equation \ref{eq:loc}].''  (1975, p. 55)  Of course, one might
contemplate an alternative theory in which the relevant complete
specification of beables included something else in addition to (or
instead of) the QM wave function; such an alternative may or may not
exhibit any nonlocal causation.  This illustrates the point made
earlier:  what we diagnose as ``nonlocal'' by applying Bell's
criterion is some particular candidate theory (as opposed to
immediately inferring, from either the empirical predictions of a
theory or a given set of empirical observations, that some non-local
causation is occuring in Nature.)  The point here is that
orthodox QM's account of the ``Einstein's Boxes'' scenario involves
non-local causation; whether any nonlocality in fact occurs in Nature
when one performs the indicated experiment involving an
$\alpha$-particle, however, is a very different question.  If, for
example, Bohmian Mechanics (rather than orthodox QM) is true, the
answer would be no.}

For ease of future reference and to fix some terminology, it will be 
helpful to lay out here a bit more explicitly the type of setup
involved in the Bell experiments, and to indicate precisely how one
gets from locality as formulated by Bell to the somewhat
different-looking mathematical condition (sometimes called
``factorizability'') from which standard derivations of Bell's
inequality proceed.  

The setup relevant to Bell's theorem involves a particle source which
emits pairs of spin-correlated particles, and two spatially separated
devices each of which allows measurement of one of several spin
components on the respective incident particle.  (In actual
experiments, the particles are typically photons with polarization
playing the role of ``spin.'')  Two experimenters, traditionally
Alice and Bob, man the two devices.  We use the symbols
$\hat{a}$ and $\hat{b}$ to refer, respectively, to the ``settings'' of
Alice's and Bob's apparatus (one usually thinks here of an axis in
space along which the polarizer or Stern-Gerlach magnetic field is
oriented), and $A$ and $B$ to refer to the
``outcomes'' of their respective spin-component measurements.
Finally, we will use the symbol $\lambda$ to refer to the ``state of
the particle pair.''  The scare-quotes around the various terms here
are an advertisement for the following discussion.

First, note that all of the symbols just introduced refer to
\emph{beables}.  There is a tendency in the literature for all of
these things (the apparatus settings, the outcomes, and the physical
state of the particle pair) to remain needlessly abstract.  But all of
these things are perfectly concrete, at least relative to some
particular candidate theory.  The setting of Alice's
apparatus, for example, refers to something like the spatial
orientation of a Stern-Gerlach device, or some sort of knob or lever
on some more black-box-ish device.  Thus, this ``setting'' ultimately
comes down to the spatial configuration of some physically real
matter, i.e., it must be reflected somehow in the beables posited by
any serious candidate theory. 
\footnote{A candidate theory which posited no
beables corresponding to such things as knobs and levers should not,
and probably could not, be taken seriously.  Bell stresses in his very
first discussion of beables that:  ``The beables must include the
settings of switches and knobs on experimental equipment ... and the
readings of instruments.'' (1975, p. 52)  For elaboration of
the sense of the term ``serious'' being used here, see Bell's (1986,
pp. 194-5).}

Likewise, the outcome of
Alice's experiment is not some ethereal event taking place in Alice's
consciousness or some other place about which serious candidate physical
theories fail to speak directly.  Rather, the outcome should be
thought of as being displayed in the post-measurement position of a
pointer (or the arrangement of some ink on a piece of paper, etc.) -- in
short, the outcome too is just a convenient way of referring to some
physically real and directly observable configuration of matter, and
so will necessarily be reflected in the beables posited by any serious
candidate theory.  The case with $\lambda$ is a little different,
because this is not something to which we have any sort of direct 
observational access.  But this only means there is significantly more
freedom about what sort of beables $\lambda$ might refer to in various
different candidate theories.

\begin{figure}[t]
\begin{center}
\includegraphics[width=3.3in,clip]{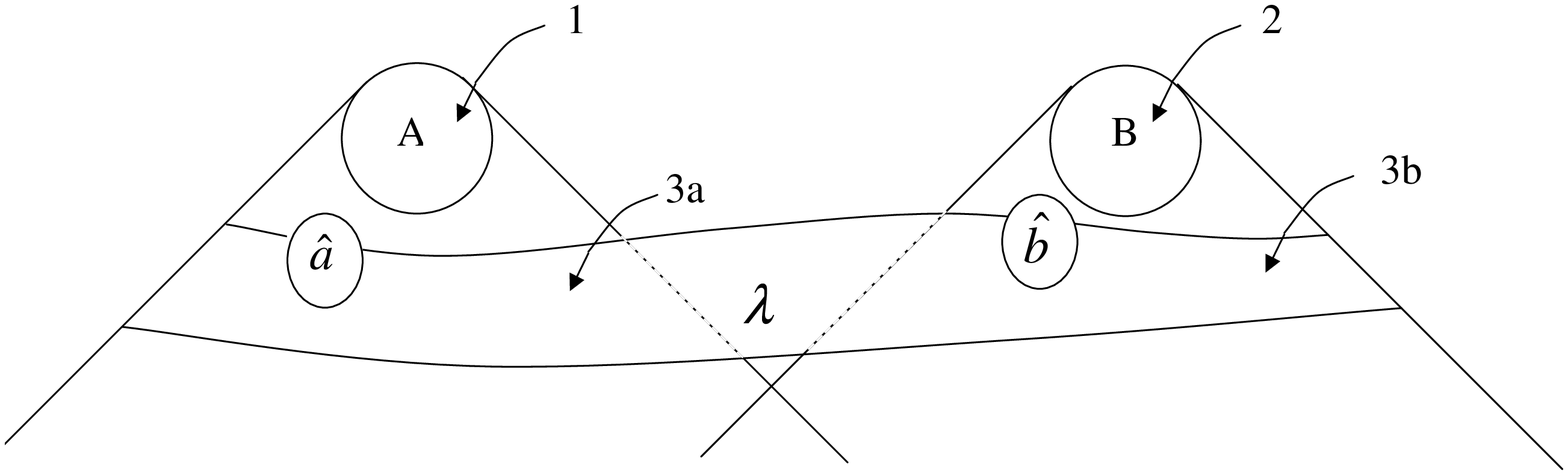}
\end{center}
\caption{
\label{fig5}
Space-time diagram illustrating the various beables of  relevance for
a discussion of factorization.  Separated observers Alice and Bob make
spin-component measurements (using apparatus characterized by
variables $\hat{a}$ and  $\hat{b}$
respectively) of a pair of entangled spin-1/2 particles.   The state
of the particles (and/or any other appropriately associated beables) is denoted
$\lambda$, and the outcomes of the two measurements are represented by
the beables $A$ (in region 1) and $B$ (in region 2).  Note that
$\lambda$ and $\hat{a}$ jointly constitute a complete
specification of beables in space-time region $3a$, which shields off
region 1 from the overlap of the past light cones of 1 and 2.
Likewise, $\lambda$ and $\hat{b}$ jointly constitute a complete
specification of beables in space-time region $3b$, which shields off
region 2 from the overlap of the past light cones of 2 and 1.   Thus
the joint specification of $\lambda$ and $\hat{a}$ will -- in a
locally causal theory -- make $\hat{b}$ and $B$ redundant for a 
theory's predictions about $A$ (and likewise, specification of
$\lambda$ and $\hat{b}$ will render $\hat{a}$ and $A$ redundant
for predictions about $B$).
}
\end{figure}

The basic space-time structure of the setup in question is sketched in
Figure 4, and the overall logic is explained in the caption.  The idea 
is simply to apply Bell's locality condition to the two measurement
outcomes $A$ and $B$ in order to assert that the probability assigned
to a given outcome (by a locally causal candidate theory) should be
independent of both the setting and outcome of the distant
experiment.  That is, in a locally causal theory, we must have
\begin{equation}
P(A | \hat{a}, \hat{b}, B, \lambda) = P(A | \hat{a}, \lambda)
\label{eq:BLA}
\end{equation}
and
\begin{equation}
P(B | \hat{a}, \hat{b}, A, \lambda) = P(B | \hat{b}, \lambda). 
\label{eq:BLB}
\end{equation}
It is then trivial to apply the definition of conditional probability
to arrive at the conclusion that the joint probability (for outcomes
$A$ and $B$) \emph{factorizes}:
\begin{equation}
P(A,B | \hat{a},\hat{b}, \lambda) = P(A|\hat{a},\lambda) \times
P(B|\hat{b},\lambda).
\label{eq:factor}
\end{equation}
Bell writes:  ``Very often such
factorizability is taken as the starting point of the analysis.  Here
we have preferred to see it not as the \emph{formulation} of `local
causality', but as a consequence thereof.'' (1990, p. 243)

It is also important that the apparatus settings $\hat{a}$ and
$\hat{b}$ are in some sense ``free.''  This is often discussed in
terms of Alice and Bob making literal last-minute free-will choices
about how to orient their devices.  What is actually required for the
proof is merely the assumption that $\hat{a}$ and $\hat{b}$ are
(stochastically)
independent of the particle pair state $\lambda$.
This comes up in the course of the derivation of Bell's inequality
when we write an expression for predicted empirical frequency of a certain
joint outcome as a weighted average over the candidate theory's
predictions for a given $\lambda$ -- that is,
\begin{equation}
E(A,B|\hat{a},\hat{b}) = \int d\lambda \, P(A,B|\hat{a},\hat{b},\lambda)
\, P(\lambda)
\end{equation}
where $P(\lambda)$ is the probability that a given particle pair state
$\lambda$ is produced by the preparation procedure used at the source.

If the probability distribution $P(\lambda)$ actually depended on
$\hat{a}$ or $\hat{b}$ -- as one would expect if the 
particle pair exerted some causal influence on the settings (or vice
versa!) or if the particle pair and the settings were mutually causally
influenced in some non-trivial way by events farther back in the past
-- then the above expression for the empirical frequency would be
invalid and the derivation of the Bell inequality wouldn't go through.
\footnote{See Bell, 1990, pp. 243-4.}

This might seem like cause for concern, especially considering that
(as displayed in Figure 4) the past light cones of $\hat{a}$ and
$\hat{b}$ overlap with the region containing
$\lambda$ -- and $\lambda$ by definition is supposed
to contain a \emph{complete} specification of beables in this region.
Given all that, one wonders how $\hat{a}$ and $\hat{b}$ could possibly
\emph{not} be causally influenced by $\lambda$ (in a locally causal
theory).  

Here we see the
reason for another aspect of Bell's carefully-phrased formulation of 
locality, which requires that beables in (the relevant) ``region 3''
(which will of course be two different regions for the two measurements)
be ``sufficiently specified, for example by a full specification...''
Here there is the implication that a
specification could be sufficient \emph{without} being complete.  For
example, some candidate theory (and this is actually true of every
serious extant candidate theory) might provide a specification of the
state of the particle pair which is sufficient in the relevant sense,
even though it leaves out some fact (say, the millionth digit of the
energy of some relic microwave background photon that happens to
fly into the detection region just prior to the measurement) which
actually exists in the relevant spacetime region.  Such a fact could
then be allowed to determine the setting $\hat{a}$ without introducing
even the slightest evidence for the problematic sort of correlation
between $\hat{a}$ and $\lambda$.  
\footnote{See Bell's discussion in his 1977, pages 100-104.}
Indeed, this is just an exaggerated
version of what happens in the actual experiments, where
carefully-isolated and independent pseudo-random-number generators are
used to produce the settings at the two stations.  

Finally, as Shimony, Horne, and Clauser have pointed out, 
\begin{quote}
``In any scientific experiment
in which two or more variables are supposed to be randomly selected,
one can always conjecture that some factor in the overlap of the
backward light cones has controlled the presumably random choices.
But, we maintain, skepticism of this sort will essentially dismiss all
results of scientific experimentation.  Unless we proceed under the
assumption that hidden conspiracies of this sort do not occur, we have
abandoned in advance the whole enterprise of discovering the laws of
nature by experimentation.   [Hence, the extra] supposition needed [to
derive Bell's inequality from Bell's locality condition] is
no stronger than one needs for experimental reasoning generically, and
nevertheless just strong enough to yield the desired inequality.''  
\footnote{A. Shimony, M.A. Horne, and J.F. Clauser, ``An Exchange on
  Local Beables,'' \emph{Dialectica}, 39 (1985) 86-110}
\end{quote}
So in the end there is really nothing worth worrying about here, i.e.,
nothing which, in the face of the experimental data conflicting with
Bell's inequalities, one might reasonably reject as an alternative to
rejecting Bell's local causality.  What
is important is that Equation \ref{eq:factor} (along with the ``freedom''
or ``no conspiracies'' assumption just discussed) entails the Bell
inequality.
The derivation is standard and will not be repeated here.  
\footnote{See, e.g., Bell 1975, pages 55-57.}
We will instead simply note
that (starting in 1975) Bell almost always referred 
to the empirically testable inequality as the ``locality inequality.''  
One can hopefully now appreciate why.

One final caveat.  There is a sense in which our
verbal description of the meanings of the relevant beables ($\hat{a}$,
$\hat{b}$, $A$, $B$, and $\lambda$) is potentially confusing.  For
example, there is no sense in which Bell makes some dubious ``extra
assumption'' that, e.g., \emph{particles} fly from the ``particle
source'' to the measurement devices.   
\footnote{Bell, 1981, p. 150}
A theory which posits no
particle beables, but only (say) waves or a wave function or \emph{whatever}
is perfectly fine and the formal locality criterion
will apply to it just the same way.  Only the words would need to
change.  Relatedly, many papers in the Bell literature raise issues
about hidden variables associated not with the particle pair, but with
the measurement devices.  Have we excluded such variables, since our
$\hat{a}$ refers only to some knob setting on Alice's device and
$\lambda$ refers only to the state of the particle pair?  No.  The
only important distinction here is that $\hat{a}$ and $\hat{b}$ refer 
to beables which are ``free'' (or ``random'') in the above sense,
while the variables $\lambda$ are somehow set by past events.

This is no doubt a fuzzy distinction, but nothing important hinges on
it.  The point is that any
microscopic features of Alice's device (hidden or not)  -- or anything
else relevant to the candidate theory's predictions for the probabilities in
question -- can be included under the ``setting'' variables ($\hat{a}$,
$\hat{b}$) \emph{or} the ``particle state'' variables ($\lambda$),
whichever seems more natural.  In short, Bell's formulation of
locality is significantly more general than might otherwise be
suggested by some of the words used to describe it.

\section{Jarrett's Analysis}

Jon Jarrett's influential analysis of Bell's locality criterion 
begins with Equations \ref{eq:BLA} and \ref{eq:BLB}, which he dubs
``strong locality.''  (Of course, we are innocuously 
changing -- and occasionally
simplifying -- Jarrett's notation to make it consistent with that
introduced above.)  For simplicity, let us focus the discussion 
on Equation \ref{eq:BLA}.  Jarrett defines two sub-conditions which,
he subsequently proves, are jointly equivalent to this ``strong locality.''

The first sub-condition Jarrett dubs ``locality.''  (It is also
sometimes referred to as ``simple locality'' or ``parameter
independence'' or ``remote context independence'' in the literature.)
\begin{equation}
P(A | \hat{a},\hat{b}, \lambda) = P(A| \hat{a}, \lambda).
\label{eq:JL}
\end{equation}
As Jarrett explains, ``Locality requires that the probability for the 
outcome
[$A$] ... be determined `locally'; i.e., that it depend only on the
state $\lambda$ ... of the two-particle system and on the state
[$\hat{a}$] of the measuring device.  In particular, that probability
must be independent of which (if any) component of spin the distant
measuring device is set to measure.''  
\footnote{Jon Jarrett, \emph{op cit.}, p. 573}

Jarrett's second sub-condition, which he dubs ``completeness'' (and is
also known as ``predictive completeness,'' ``outcome
independence,'' ``remote outcome independence'' and ``conditional 
outcome independence'') is the following:  
\begin{equation}
P(A | \hat{a}, \hat{b}, B, \lambda) = P(A | \hat{a}, \hat{b}, \lambda)
\label{eq:JC}
\end{equation}
which ``asserts the stochastic independence of
the two outcomes in each pair of spin measurements.''
\footnote{\emph{Ibid.}, p. 578}

It is easy to see that, indeed, ``locality'' and ``completeness'' are
jointly equivalent to Bell's locality condition as expressed in
Equation \ref{eq:BLA}.  First, $P(A|\hat{a},\hat{b},B,\lambda)$ is,
under the assumption of ``completeness,'' equal to
$P(A|\hat{a},\hat{b}, \lambda)$.  And this, in turn, is equal to $P(A |
\hat{a}, \lambda)$ under the assumption of ``locality''.  So
``locality'' and ``completeness'' jointly entail ``strong locality.''
And likewise ``strong locality'' clearly entails both ``locality''
and ``completeness.''  So the two sub-conditions are, indeed,
equivalent to Equation \ref{eq:BLA}.  

The alleged significance of this decomposition, however, emerges
only from Jarrett's discussion of the physical interpretation of his
two sub-conditions.  

First, Jarrett argues that ``locality'' is equivalent to the
prohibition of superluminal signaling, and 
hence expresses just what relativity requires of other theories.  As
was mentioned in the introduction, both steps of this argument have
been found wanting.  First, it is only in combination with some assumptions
about the \emph{controllability} of various beables (notably
$\lambda$) that Jarrett's ``locality'' is equivalent to the
prohibition on superluminal signaling.  There is at least one
extant, empirically viable theory (Bohmian Mechanics) which violates
Jarrett's ``locality'' condition and yet doesn't permit the
possibility of superluminal signaling, precisely because the relevant
states cannot (as a matter of principle, as predicted by the
theory) be sufficiently controlled.  And second, it is dubious to
claim that the prohibition of superluminal \emph{signals} adequately
captures relativity's fundamental speed limit.  This would, for
example, render Bohmian Mechanics consistent with SR despite its need
to postulate a dynamically privileged reference frame -- a ``gross
violation of relativistic causality'' according to Bell. (1984, p. 171)
But since these
problems have been discussed elsewhere in the literature, we leave
them aside here and focus instead on Jarrett's physical interpretation
of his second sub-condition, ``completeness.''  

Jarrett elaborates the meaning of his ``completeness'' condition with
the following example:
\begin{quote}
``A simple (and incorrect) model for the Bell-type correlated spin
phenomena may serve as a useful illustration.  Suppose, purely for the
sake of illustration, that spin is correctly represented as an
ordinary classical angular momentum.  Suppose further that when a pair
of particles is prepared in the singlet state, the spin vectors for
the two particles are aligned exactly anti-parallel to each other.
Moreover, given an ensemble of such two-particle systems, suppose that
each direction in space is equally likely to be the direction of
alignment for an arbitrarily selected member of the ensemble.
Finally, if the unit vector [$\hat{a}$] gives the direction along
which the
axis of the Stern-Gerlach apparatus is aligned ... and if [$\hat{s}$]
is the spin vector of the particle ... then the outcome of that
measurement is $+1$ if [$\hat{a}\cdot \hat{s} > 0$] and $-1$ if
[$\hat{a}\cdot \hat{s} < 0$].  
\end{quote}
Jarrett then makes what amounts to the following point:  for this
model, we clearly have that Alice's particle (from a randomly selected
member of the ensemble) is equally likely
to be found in the $A = +1$ and $A = -1$ states along any arbitrary
direction $\hat{a}$.  Thus, for example,
\begin{equation}
P (A \! = \! +1 | \hat{a}, \hat{b}, \lambda) = \frac{1}{2}
\label{eq:comp1}
\end{equation}
where $\lambda$ is the singlet state -- evidently meaning, in the context of
this example, the state description according to which the particle has
some definite but completely unknown spin direction $\hat{s}$.  

On the other hand, it is built into the model that, for $\hat{a} =
\hat{b}$, the outcomes of Alice's and Bob's measurements will be
perfectly anti-correlated.  Hence, if we additionally specify the
outcome of Bob's experiment, the outcome of Alice's is \emph{fixed}.
Suppose, for example, that $B = +1$.  Then $A = +1$ is forbidden,
i.e., 
\begin{equation}
P (A \! = \! +1  | \hat{a}, \hat{b}, B, \lambda) = 0. 
\label{eq:comp2}
\end{equation}
Comparing the previous two equations, we see that ``completeness is
clearly violated.''  \footnote{\emph{Ibid.}, p. 580}

Jarrett elaborates: 
\begin{quote}
``The probabilities specified for this model are
grounded in a blatantly incomplete description of the two-particle
state.  In the context of this model, if the theory assigns
probabilities only on the basis of the occupancy by the two-particle 
system of the singlet state, then conditioning on the outcome [$B$] of a
[$\hat{b}$]-component spin measurement on [Bob's particle] may well
yield a different probability for the outcome of a spin measurement on
[Alice's particle] than would have been given by the corresponding
unconditioned probability (i.e., 1/2).  This is so because, if the
outcome of the measurement on [Bob's particle] is $+1$ [and if
$\hat{a} = \hat{b}$], then it may be
inferred that [$\hat{a} \cdot \hat{s}_A < 0 $] (with probability 1),
where [$\hat{s}_A$] is the spin of Alice's particle....
The outcome of the measurement on [Bob's particle thus] provides
information about [Alice's particle] which was not included in the
incomplete state description [$\lambda$].''  \footnote{\emph{Ibid.},
  p. 580}
\end{quote}
The important conclusion is this:
the fact that the probability assigned to a certain outcome
for Alice's experiment 
depends, in violation of Equation \ref{eq:JC}, on the
outcome of Bob's experiment, does not mean that there is any
relativistically-forbidden superluminal \emph{causal influence} (e.g.,
from Bob's outcome to Alice's).  
That is the whole point of the illustrative example, in
which (by assumption) the outcome of Alice's experiment is determined
exclusively by factors (namely $\hat{a}$ and $\hat{s}_A$) which are
present at her location.  No nonlocal causal influence exists.  
Instead, the violation of Equation \ref{eq:JC} indicates only that we
were dealing with \emph{incomplete state descriptions}, such that
Bob's outcome provides some \emph{information} (usefully supplementing
what was already contained in $\lambda$) which warrants an updating of
probabilities.

On the basis of this example, 
Jarrett thus urges the following physical interpretation of
his two sub-conditions:  a violation of ``locality'' would allow the
possibility of sending superluminal signals, and hence indicates
clearly the existence of some relativity-violating superluminal causal
influences; on the other hand, a violation of ``completeness'' does
not indicate the existence of any relativity-violating influences, but
instead suggests only that the state descriptions of the theory in
question are not complete.  It is clear that these physical
interpretations of the two conditions were the basis for Jarrett's
decision to name them ``locality'' and ``completeness'' respectively.  

Here is Jarrett's summary of the cash value of this decomposition
\emph{vis-a-vis} Bell's theorem and the associated experiments:  these
together provide very strong evidence ``that
strong locality cannot be satisfied by any empirically adequate
theory.  Since locality is contravened only on pain of a serious
conflict with relativity theory (which is extraordinarily
well-confirmed independently), it is appropriate to assign the blame
to the completeness condition.  ...[O]ne must conclude that certain
phenomena simply cannot be adequately represented by any theory which
ascribes properties to the entities it posits in such a way that no
measurement on the system may yield information which is both
non-redundant (not deducible from the state descriptions) and
predictively relevant for distant measurements.  That `information' is
not (neither explicitly nor implicitly) contained in the `incomplete'
state description.''  \footnote{\emph{Ibid.}, p. 585}

\section{Comparison}

The fundamental origin of the disagreement between Bell and Jarrett
should now be clear:  the two authors do not understand (e.g.)
Equation \ref{eq:BLA} in the same way.  For Bell, the variables
$\lambda$ in this formula (together with $\hat{a}$ as per the previous 
discussion) constitute a \emph{complete} (or perhaps merely
sufficient) specification of beables in some
space-time region that has the same relation to Alice's experiment  
that region 3 (of Figure 1) had to region 1.  Jarrett, by contrast,
is agnostic about the completeness of the description afforded by
$\lambda$.  

Strictly speaking,
therefore, Jarrett's decomposition of Equation \ref{eq:BLA}
is not a decomposition of
Bell's locality condition, but, rather, a decomposition of some sort of
no-correlation condition 
\begin{equation}
P(b_1 | \bar{B}_3, b_2) = P(b_1 | \bar{B}_3)
\label{eq:nocorr}
\end{equation}
which is analogous to Equation \ref{eq:loc}, except that, following
the notation of Section II, the variables $\bar{B}_3$ are \emph{not} 
assumed to provide a complete specification of
beables in the relevant spacetime region.  But as we have already
discussed in Section II, and as Bell was perfectly aware,
a violation of Equation \ref{eq:nocorr} does
not necessarily indicate the presence of any nonlocal causation in the
candidate theory in question.  Indeed, it is precisely to close off the
avenue eventually taken by Jarrett (that is, blaming a violation of this
``no-correlation'' condition on the incompleteness of the specified
beables) that Bell specifically stresses the importance that ``events
in [the relevant region] be specified completely.'' (1990, p. 240)

This confusion -- this departure from Bell's
actual locality criterion -- is the ultimate basis for Jarrett's
choice of terminology, and also for any initial plausibility of his
project of establishing ``Peaceful Coexistence'' by showing that a
violation of the locality criterion needed for Bell's inequality (in
particular, a violation of his ``completeness'' sub-condition) need
not indicate any conflict with relativistic local causality. 

Of course, one could simply apply Jarrett's decomposition strategy to
Bell's actual locality condition.  That is, it is true that, as
Jarrett claimed to have shown, Bell's locality condition can be
decomposed into two Jarrett-like sub-conditions -- namely, our
Equations \ref{eq:JL} and \ref{eq:JC} but now with the requirement
(inherited from Bell's actual locality condition) that $\lambda$ 
provide a complete (or sufficient) specification of relevant beables
(as posited by some candidate theory whose locality is being assessed
by the locality condition).  
But \emph{this} decomposition fails to have any of the physical
implications urged by Jarrett.  In particular, a violation of the
(strengthened) ``completeness'' condition 
\begin{equation}
P(A|\hat{a},\hat{b},B,\lambda) = P(A|\hat{a},\hat{b},\lambda)
\label{eq:BJcomp2}
\end{equation}
(formally equivalent to Jarrett's ``completeness'' sub-condition, but now,
with Bell but contra Jarrett, with the insistence that $\lambda$ and 
$\hat{a}$ jointly provide a complete description of beables in some
spacetime region through the past of $A$ which divides $A$ off
completely from the overlap of the past light cones of $A$ and $B$)
has absolutely nothing to do with the completeness of state descriptions,
but instead indicates the
presence of some nonlocal causation (in violation of the causal
structure outlined in Figure 1) in the candidate theory in question.

The most that could be said to distinguish the two sub-conditions is that,
since $\hat{b}$ is (by definition) controllable and $B$ (most likely) 
isn't, a violation of \ref{eq:JL} is (all other things being equal) 
more likely to yield the possibility of superluminal \emph{signaling}
than a violation of \ref{eq:JC}.  But that only matters if we drop
what Bell calls ``fundamental relativity'' and instead read SR
instrumentally, as prohibiting superluminal signalling but allowing in
principle superluminal causation (so long as it can't be harnessed by
humans to transmit messages).   That is, at best, a dubious and 
controversial reading of SR, as already mentioned.  \footnote{See Bell's
discussions of ``controllable'' beables and causation \emph{vs.} 
signaling in his 1975, pp. 60-61; 1984, p. 171; and 1990, pp. 244-246.}

It is sometimes suggested 
\footnote{See, e.g., p. 153 of Harvey Brown's half of ``Nonlocality in Quantum
  Mechnics,''  Michael Redhead and Harvey Brown, \emph{Proceedings of
    the Aristotelian Society, Supplementary Volumes}, Vol. 65 (1991),
  pp. 119-159.}
that the relation between Bell and Jarrett
on this point is one of basic agreement, since both held that the
validity of Equation \ref{eq:JC}
has something to do with the completeness of the physical theory in
question.  This is doubly incorrect.  First, the mathematical
condition in question is, for Jarrett, a \emph{formulation} of
completeness; it is, for Bell, (part of) a formulation of local
causality which functions appropriately as such \emph{only if the 
relevant symbols in the formula stand for a complete specification 
of the relevant beables}.  According to Jarrett, Equation \ref{eq:JC} is
supposed to tell us whether completeness holds.  For Bell, on the
other hand, the equation (correctly understood) already presupposes
that completeness holds.  (If it doesn't hold, the condition is useless for
his purposes, and states only that the two outcomes are uncorrelated.)  
And second, where Jarrett (and most subsequent commentators)
regard the ``completeness'' in question as a property of
\emph{theories}, Bell regards his ``completeness'' as a property of a
certain \emph{specification of beables} relative to some candidate
theory.  In this context, the separate question of whether the theory
itself is complete (i.e., whether its posited beables capture
everything that really exists) simply doesn't come up.

Note also that, even on his own terms, i.e., even leaving aside his 
departure from Bell's actual
locality criterion, Jarrett's formulation of completeness actually
fails as a criterion for assessing the completeness of the relevant
physical state descriptions.

Consider again Equation \ref{eq:JC}
where for the moment we follow Jarrett and interpret $\lambda$ here as
providing some kind of state description, but not necessarily a
complete one.  Jarrett has shown (with the example discussed
in Section III) that a violation of this condition can sometimes be blamed
on the use of incomplete state descriptions, with no implication of
any superluminal causation.  But it is equally easy to display an
example in which violation of Jarrett's condition \emph{cannot} be 
blamed on incomplete state descriptions, but instead indicates the
presence of superluminal causation.

Consider a toy model discussed by Maudlin, in which each particle in the
pair is indeterminate (in regard to its spin along any particular
direction) until one of the particles encounters a spin-measurement
apparatus; at this point, this ``first'' particle flips a coin to
decide whether to emerge from the $+1$ or the $-1$ port of the
apparatus, and sends an instantaneous tachyon signal to the other
particle in the pair, instructing it as to how it should behave in
order to give rise to the correct quantum mechanical joint outcomes.
\footnote{See Maudlin, \emph{op cit.}, pg 82.  After presenting this model as
  a simple example of how the observed spin correlations might
  arise, Maudlin uses it to counterexample the claim, made by Don
  Howard and others, that Jarrett's ``completeness'' and ``locality''
  can be mapped, respectively, onto the ``separability'' and
  ``locality'' conditions which emerge from some of Einstein's
  comments about local causality.  As Maudlin points out, the toy
  model is perfectly separable in the sense of Einstein, and yet
  violates what Howard \emph{et al.} would have us take as a mathematical
  formulation of Einstein's ``separability'' (namely, Jarrett's
  ``completeness.'')  Curiously, however, Maudlin does not
  mention this model in his (earlier) discussion of Jarrett.}
Assuming, as before, that $\hat{a} = \hat{b}$, and that Bob's particle
arrives at its detector first
\footnote{Of course, since Alice's and Bob's measurements are, by
  hypothesis, space-like separated, there is no relativistically
  unambigous meaning to ``first.''  But that is really the whole
  point.  This model explicitly involves anti-relativistic
  superluminal causation, so part of the model is that relativity is
  wrong and there exists some dynamically privileged reference frame
  which gives an unambiguous meaning to this ``first'' (and also to
  the ``instantaneous'' in the description of the tachyon signal).}
with the result $B = +1$, 
this model makes the same predictions, Equations \ref{eq:comp1} and
\ref{eq:comp2}, as the local deterministic model discussed by
Jarrett.  And so this model, like Jarrett's, violates Jarrett's
``completeness'' condition.  And yet clearly with this model there is
no blaming that violation on the use of incomplete state descriptions
-- instead here it is obvious (by construction) that the violation is
due to the presence of superluminal causal influences (in the
particular form of ``tachyon signals'').    

This should not be surprising.  We have already argued that, if one
follows Bell in requiring $\lambda$ to constitute a complete state
description, then a violation of Jarrett's ``completeness'' can \emph{only} be
understood as indicating the presence of nonlocal causation.  The
point here is that, even if we follow Jarrett in remaining agnostic
about the completeness of the description afforded by $\lambda$, we
cannot \emph{necessarily} say that a violation of ``completeness'' is
compatible with relativity's prohibition on superluminal causation.
It might be (as shown by Jarrett's model).  But it might not be (as
shown by Maudlin's).  

The correct conclusion is therefore as follows:  a violation of
Jarrett's ``completeness'' condition (where we are openly agnostic
about the completeness of the state description $\lambda$) means
\emph{either} that we have relativity-violating nonlocal
causation, \emph{or} that we were dealing with incomplete state 
descriptions.  And \emph{this} dilemma is precisely that posed already in
1935 by Einstein, Podolsky, and Rosen:  \emph{either} we concede Bohr's claim
that the QM description of states is complete and accept the reality
of the nonlocal causation present in that theory; \emph{or} we reject the
completeness claim and adopt a different (``hidden variables'') theory
which (the authors thought) could restore locality.
\footnote{A. Einstein, B. Podolsky, and N. Rosen, ``Can
  quantum-mechanical description of physical reality be considered 
  complete?'' \emph{Phys. Rev.} 47 (1935), 777-780.  For some more
  recent discussion, see T. Norsen, ``Einstein's Boxes,'' \emph{op
    cit.} and references therein.}

Jarrett also sees a similarity between his conclusion and that of EPR
(namely that QM is not complete).  But where EPR considered this a
defect to be corrected in some alternative ``hidden variable'' theory
(which they hoped would also restore local causality), Jarrett argues
that incompleteness is not a defect of orthodox QM, but, rather, a
fact of nature:  
\begin{quote}
``By separating out the relativistic component of the
strong locality condition ... there emerges a clarification of that
class of theories excluded by the Bell arguments:  the class of
theories which satisfy completeness.  Although the term
`incompleteness' may connote a defect (as if, as was the case for the
model discussed [above], all incomplete theories may be
`complet\emph{ed}'), incomplete theories (e.g., quantum mechanics) are
by no means \emph{ipso facto} defective.  On the contrary, when the
result of Bell-type experiments are taken into account, the truly
remarkable implication of Bell's Theorem is that incompleteness, in
some sense, is a genuine feature of the world itself.''
\footnote{\emph{Op cit.}, p. 585}
\end{quote}
This sort of claim is echoed also in Jarrett's later writings. 
\footnote{L.E. Ballentine and Jon P. Jarrett, 
``Bell's theorem:  Does quantum mechanics contradict relativity?'' 
\emph{Am.J.Phys.} 55(8) (August 1987), 696-701; Jon Jarrett, ``Bell's
Theorem:  A Guide to the Implications'' in J. Cushing and E. McMullin,
eds., \emph{op cit.}}

Ballentine and Jarrett also give an argument that the ``completeness''
of interest to EPR is a stronger form of Jarrett's ``completeness'' in
the sense that the former entails the latter.  (This occurs in the
context of their agreeing with EPR that QM is incomplete, which they
establish by arguing that, since the correct conclusion from
Bell/experiment is that Jarrett's ``completeness'' condition fails,
then the stronger EPR completeness condition must also fail, just as the
EPR authors claimed.)  But this association is mistaken, since
the argument Ballentine and Jarrett display sneaks in the additional
premise of (EPR's version of) local causality.  \footnote{``But this
  prediction was made without in any way disturbing particle $L$,
  since the $R$ device is at spacelike separation from it...''  This
  mistake was also noted by Andrew Elby, Harvey R. Brown, and Sara
  Foster in ``What Makes a Theory Physically Complete?''
  \emph{Found. Phys.} 23(7),  971-985 (1993).  Note
  also that this error leads Ballentine and Jarrett to remark, in
  passing, that EPR's concept of completeness entails determinism all
  by itself, which is surely a misrepresentation of the worries of
  Einstein and his followers.  See Section 1 of Bell's 1981,
  pp. 139-145, and the important footnote 10 -- already partially
  quoted in the introduction -- on p. 157.}
And this is the premise that does (literally) 
all the work in the EPR argument.  

There is thus no apparent sense at all in which EPR's completeness
has anything to do with Jarrett's, except that it was precisely in
the face of QM's violation of Jarrett's ``completeness'' that EPR
argued (correctly) that QM was a non-local theory which, perhaps, could be
replaced by a locally causal alternative theory by adding hidden
variables (or jettisoning the description in terms of wave functions
entirely).  Indeed, at the end of the day, it is pretty clear that 
Jarrett's condition has nothing to do with the completeness of physical
state descriptions.  In concluding (from Bell's Theorem and the
associated experiments) that reality itself is ``in some sense''
incomplete, Jarrett makes clear that he is no longer using the term
with anything like its ordinary meaning -- namely, a description which 
leaves nothing (relevant) out.  By what standard, exactly, could
``the world itself'' be supposed to have left something out?  

It is worth stepping back, therefore, and clarifying what, if 
anything, one can say about ``completeness.''
There are two related senses on the table.  First of all, a theory may
be said to be ``complete'' or ``incomplete'' in relation to external
physical reality.  In this sense, a theory is complete if and only if
it captures or describes everything (relevant) that in fact really
exists.  This is of course just the sense of completeness of interest
to EPR.  Their argument, in essence, was that relativistic local
causality (which they simply took for granted) combined with certain
empirical predictions of QM entailed the existence
of some ``elements of reality'' which had no counterpart in the QM
description of reality.  That is, the QM description left something out;
it was hence
an incomplete theory.  

Since this usage of ``completeness'' involves a comparison between
theories and external reality (to which our best access is precisely
through theories!), there is a tendency for it to be regarded as
``metaphysical''.  
Perhaps this apparently metaphysical flavor is
responsible for Jarrett's suppression of
Bell's requirement that $\lambda$ contain a complete specification of
the relevant beables.  Since there'd be no way to verify whether a given
specification was or wasn't complete in this sense, one might think,
Bell's requirement is meaningless and might as well just be dropped.  

But this attitude fails to appreciate one of 
Bell's important advances -- namely, that his formulation of local
causality is a criterion for assessing the locality of \emph{candidate
  theories}.  As already discussed in Section II, Bell's ``complete
specification of beables'' simply does not mean a specification that
captures everything which in fact really exists; rather, it means a
specification which captures everything which is \emph{posited} to
exist \emph{by some candidate theory}.  There is thus nothing the
least bit metaphysical or obscure about Bell's requirement.  For any
unambiguously formulated candidate theory, there should be no question 
about what
is being posited to exist.  And so there will be no ambiguity about
what a complete description of relevant beables should consist of, and
hence no ambiguity about the status -- \emph{vis-a-vis} local
causality  -- of a given well-formulated theory.  

There will of course still be
difficult questions about how to decide whether a given candidate
theory is \emph{true}, and hence whether the particular \emph{sort} of
non-local causation contained in it accurately describes some aspect
of Nature.  But the miracle of Bell's argument is that we
need not know which theory is true, in order to know that the true
theory (whatever it turns out to be) will have to exhibit non-local,
super-luminal causation.  There is thus no escaping Bell's conclusion
that some sort of non-local causation (in violation of the structure
displayed in Figure 1) exists in Nature -- in apparent conflict with
what most physicists take to be the requirements of SR.

\section{Discussion}

In the previous section, we interpreted Jarrett as claiming that a
violation of his ``completeness'' condition in no way implied the
presence of non-local causation, but instead only implied that the
state descriptions used in the test had been incomplete.  This is both
fair and unfair -- fair because Jarrett does hang his entire case for
the plausibility of his terminology and his physical interpretation of
the two sub-conditions on precisely this view, but also unfair because
Jarrett also later seems to acknowledge that, ultimately, his
``completeness'' condition has to be understood very differently.  For
example, he remarks in a footnote that ``completeness, too, has the
character of a `locality' condition.''  \footnote{Jon Jarrett,
  \emph{op cit.}, p. 589}  And the trend in the Bell
literature since Jarrett's paper has certainly been to concede that a
violation of Jarrett's ``completeness'' cannot be quite so trivially
written off (as involving a mere updating of information in the face
of having previously used incomplete state descriptions), but rather
must be understood as indicating \emph{some} sort of non-locality or
``holism'' or ``non-separability'' or non-causal ``passion at a
distance.''  

We do not, therefore, wish to claim that Jarrett (and those who follow
him in thinking his decomposition is in some way or other helpful in
understanding Bell's locality condition and/or in establishing the
peaceful coexistence of SR and QM  \footnote{For a systematic
  review of the recent literature, see Berkovitz, Joseph, "Action at a
  Distance in Quantum Mechanics", The Stanford Encyclopedia of
  Philosophy (Spring 2007 Edition), Edward N. Zalta (ed.), URL =
  http://plato.stanford.edu/archives/spr2007/entries/qm-action-distance/}
) fully commits precisely the mistake presented (through the stark
contrast to Bell's views) in the previous section.  Rather, we intend
only the weaker claim that Jarrett \emph{et al.} have been led, by
Jarrett's initial analysis, down a path which is \emph{obviously
  untenable} once one clearly understands Bell's own formulation of
local causality (including especially the parallel status -- namely,
both are \emph{beables} -- of ``settings'' and ``outcomes,'' and the 
crucial distinction between superluminal causation and superluminal 
signaling).

It turns out to be a rather subtle question whether or not SR
genuinely requires local causality in the sense of Figure 1.  
\footnote{Much of Maudlin's excellent book, \emph{op cit.}, is
  dedicated to exploring just this question.}
But if one grants this (and virtually all physicists and commentators
\emph{do}), then it really is possible to establish an 
\begin{quote}
``essential
conflict between any sharp formulation [of QM] and fundamental
relativity.  That is to say, we have an apparent incompatibility, at
the deepest level, between the two fundamental pillars of contemporary
theory...'' (Bell, 1984, p. 172)
\end{quote}
The widespread claims to the contrary -- i.e., the claims that instead
Bell's theorem refutes only some already-dubious, dogmatic,
philosophically-motivated program to restore ``determinism'' or
``classicality'' or ``realism'' (and I mean here both classes of such
claims mentioned in the introduction) -- turn out inevitably to have
their roots in a failure to appreciate some aspect of Bell's own
arguments.  

There is, in particular, a tendency for a relatively 
superficial focus on the relatively formal aspects of Bell's
arguments, to lead commentators astray.  For example, how many
commentators have too-quickly breezed through the prosaic first
section of Bell's 1964 paper (p. 14-21) -- where his reliance on the EPR
argument ``\emph{from locality to} deterministic hidden variables''
is made clear -- and simply jumped ahead to section 2's 
Equation 1 (p. 15), hence erroneously inferring (and subsequently
reporting to other physicists and ultimately teaching to students) 
that the derivation ``begins with deterministic hidden variables''?  
(1981, p. 157)  Likewise, we have here explored in detail 
a similar case of too-quickly accepting some formal version of a
premise used in Bell's derivation (such as ``factorizability'') 
while failing to appreciate the rich conceptual context that gives it
the precise meaning Bell intended.  

Our final conclusion, therefore, is a plea -- directed at physicists
in general, but commentators on Bell's theorem, textbook writers, and
students in particular -- to simply read (and not just read, but
\emph{read}) Bell's writings.  They are
truly a model of clarity and physical insight, and almost always
convey the essential ideas much more lucidly and tersely than anything
in the secondary Bell literature.  (I have no doubt this applies even to
the current essay!)  
Bell himself, in the preface to the first edition
of his compiled papers (\emph{Speakable and Unspeakable in Quantum
  Mechanics}) suggests that ``even quantum experts might begin with
[chapter] 16, `Bertlmann's socks and the nature of reality', not
skipping the slightly more technical material at the end.''  It is
hard to disagree with that advice, although a strong case could be
made also for Bell's 1990 essay (written after the first edition of
the book, and hence included only in the more recent second edition)
`La nouvelle cuisine,' in which the central importance and meaning of
``local causality'' is emphasized in lucid detail.  

If more physicists would only study Bell's papers instead of relying 
on dubious secondary reports, they would, I think, come to appreciate
that there really is here a serious inconsistency to worry about.  A
much higher-level inconsistency between quantum theory 
and (general) relativity
has been the impetus, in recent decades, for enormous efforts spent
pursuing (what Bell once referred to as) ``presently fashionable
`string theories' of `everything'.''  (1990, p. 235)  How might
a resolution of the more basic inconsistency identified by Bell shed
light on (or radically alter the motivation and context for) attempts 
to quantize gravity?  We can't possibly know until (perhaps long
after) we face up squarely to Bell's important insights.

\end{document}